\documentclass[aps,pra,superscriptaddress,twocolumn]{revtex4}
\usepackage{amsmath,amsthm}
\usepackage{graphicx}
\usepackage{amssymb}
\usepackage{bbm, color}
\usepackage{epstopdf}
\usepackage{textcomp}

\begin{document}

\title{Measurement-device-independent and arbitrarily loss-tolerant verification of quantum steering}

\author{InU Jeon}
\author{Hyunseok Jeong}
\affiliation{Center for Macroscopic Quantum Control, Department of Physics and Astronomy,
Seoul National University, Seoul 08826, Korea}

\date{\today}

\begin{abstract}
  We propose a method to verify quantum steering for two qubit states with an arbitrary amount of null results when both the steering and steered parties cannot be trusted. We converted the steering inequality proposed in a recent article [Phys.~Rev.~X {\bf2}, 031003 (2012)] to a corresponding measurement-device-independent steering criterion that depends on the heralding efficiency of the steering party, number of measurement settings, and imperfection of the state preparation. As a result, for a relative frequency of valid measurement outcomes $\eta_\textsc{h}$, the steering can be verified using a number of different measurement settings larger than $1/\eta_\textsc{h}$ and maximally entangled states. Furthermore, steerability is guaranteed as long as the measurement efficiency of the steered party is non-zero. Our result is useful for loss-tolerant and measurement-device-independent steering tasks.

\end{abstract}

\maketitle

\section{Introduction}

  Nonlocal correlations between distant parties are one of the most interesting features of quantum mechanics. One may classify nonlocal correlations into three categories: Bell-nonlocality, steerability, and entanglement~\cite{Wiseman07, Jones07}. Bell-nonlocality is the strongest nonclassical property among the three that rejects any explanation of phenomena based on local realistic theory~\cite{Bell66}. It enables various quantum information tasks such as secure communication protocol~\cite{Hebert75, Ekert91, Acin06, Acin07}, random number generation~\cite{Colbeck07, Pironio10, Colbeck11}, and self-testing~\cite{Popescu92, Mayers04, McKague12, Reichardt13}. However, while Bell nonlocal states are very useful in non-trivial verifications of quantum operations, their implementation is highly demanding.

  Entanglement may be understood as a nonlocal correlation monotonic under local operations and classical communications. Although the requirement for a quantum state to be entangled is less stringent than that for Bell-nonlocality, entanglement does not always guarantee a successful test of Bell-nonlocality. In this respect, steerability, as an intermediate property, is less demanding to implement for a reasonable scope of applications compared with Bell-nonlocality. Steerability is a type of nonlocality that rejects explanation of phenomena based on a \textit{combination} of any local realistic theory and local quantum model. By definition, it is an intermediate nonlocality between entanglement and Bell-nonlocality, and thus has a wider range of applicability than entanglement and is less complicated to implement than Bell-nonlocality. Since its introduction ~\cite{Wiseman07}, many theoretical developments~\cite{Chen13, Wang14, Kiukas17, Reid89, Wiseman07, Jones07, Cavalcanti13, Bennet12, Hall16, Schneeloch13, Nagy16, Cavalcanti09, Saunders10, Kocsis15, He13, Karthik15, Reid13, Jones11, Uola15, Evans13, Vallone13, Evans14, Skrzypczyk15} and experimental verifications~\cite{Smith12, Kocsis15, Bennet12, Sun14, Saunders10, Sun16, Bartkiewicz16, Weston18, Wittmann12, Armstring15, Handchen12} have been studied, and their practical applications are still being investigated~\cite{Bartkiewicz16, Branciard12, Piani15, Passaro15, He15, Supic16, Gheorghiu17}. A good and comprehensive review on quantum steering is recently published.~\cite{Cavalcanti17}

  One of the advantages of steerability is its loss-tolerant property~\cite{Bennet12, Evans13, Evans14, Vallone13, Skrzypczyk15}, which permits the verification of steering despite an arbitrarily low measurement success rate. When the steering party is not allowed to report null results, a loss-tolerant steering criterion for arbitrary pure entangled two qubit states was proposed~\cite{Vallone13}. For the case that the steering party is allowed to report null results, a loss-tolerant steering criterion was obtained for maximally entangled two qubit states~\cite{Bennet12} and arbitrary pure entangled two qudit states~\cite{Skrzypczyk15}. These findings can close the detection loophole in the steering verification process under an arbitrary amount of detectable errors on the steering party. However, they are proposed under the one-sided-device-independent (1s-DI) scenario --- the measurement outcomes reported by the steered party must be correct. Thus, if the steered party reports false outcomes or their measurement apparatus is imperfect, this assumption is not valid, which in turn yields a loophole in the determination of steerability.

  To overcome the difficulty, a more elaborate steering scheme, which is independent of measurement devices of the steering party (say Alice) and the steered party (say Bob), was proposed and shown to be equivalent to the 1s-DI scenario~\cite{Cavalcanti13}, and its experimental verification has been demonstrated~\cite{Kocsis15}. This scheme is called a quantum refereed steering (QRS) game, and it replaces the assumption of trustfulness on Bob by sending information-encoded quantum states to him. The QRS game, however, is not fully device-independent because it assumes perfect preparation of the quantum states provided to Bob, or it requires tomography on the provided quantum states. Therefore, the QRS game is at best measurement-device-independent(MDI). Nonetheless, QRS still supplies more reliable steering verification because the generating device for state preparation or the measurement device for tomography is open to a test by an external party. In the QRS game, however, unlike the 1s-DI steering scenario, no scheme to overcome an arbitrary amount of detectable errors has been proposed yet.

  In this paper, we show that verification of steering in the QRS game is possible with arbitrary measurement efficiencies of both parties, when one-way communication from Bob to Alice is allowed. To this end, we first analyze how the 1s-DI steering inequality can be converted to that in the QRS game, and discuss some appropriate methods to deal with losses. Subsequently, we convert the arbitrarily loss-tolerant steering inequality in the 1s-DI scenario~\cite{Bennet12} to a corresponding steering criterion in the QRS game, known as the score function, in the canonical way~\cite{Kocsis15}. We shall show that the score function is indeed a steering criterion even when we allow one-way communication from Bob to Alice. Finally, we show that the effect of measurement efficiency of the steered party does not affect steerability, thus concluding that our steering criterion is arbitrarily loss-tolerant without trust on both parties.

\section{Steering and QRS game}

  The general steering protocol can be concisely summarized by Fig.1. On this protocol, on may define the QRS game introduced in Ref.~\cite{Cavalcanti13} as follows.

  \begin{enumerate}
    \item Preparation Stage --- The referee prepares sets of information $\{j\}=\mathcal{J}$, $\{s\}=\mathcal{S}$ with some probability distributions $p(j)$ and $q(s)$, and encodes $s$ in linearly dependent quantum states $\omega_{j,s}$. The referee also sets a payoff $\mathcal{P}(a,b,j,s)$ that Alice and Bob gain when Alice reports $a$ and Bob reports $b$, given that Alice receives $j$ and Bob receives $s$. Subsequently, the sum of payoff, or score, that Alice and Bob will gain in the game is given by
    \begin{align}
      S = \sum_{j,s,a,b} p(j)q(s)\mathcal{P}(a,b,j,s)P(a,b|j,s),
    \end{align}
        where $S$ denotes the score, $P(a,b|j,s)$ is a conditional probability that Alice and Bob yield $a$ and $b$ when they receive $j$ and $s$, respectively. Alice and Bob can share some quantum state beforehand and their goal is to maximize their score using the shared state. Therefore, Alice and Bob establish optimal strategy to maximize the score.

    \item Verification Stage --- The game starts when the referee provides information $j \in \mathcal{J}$ to Alice and information encoded quantum states $\omega_{j,s}$ to Bob. Once the game started, no more communication from Alice to Bob is allowed while from Bob to Alice may be permitted. Based on the information $j$ and quantum state $\omega_{j,s}$, Alice and Bob choose $a$ and $b$ according to their optimal strategy and thereafter send them back to the referee. They obtain payoff $\mathcal{P}(a,b,j,s)$ for each round of the game. After repeating sufficiently many rounds, the highest score achievable using unsteerable states is determined, and is called as the steering bound. Therefore, violation of the steering bound implies a positive verification of steering, and the corresponding shared state is steerable.
  \end{enumerate}

\begin{figure}[t]
\includegraphics[width=8.0cm]{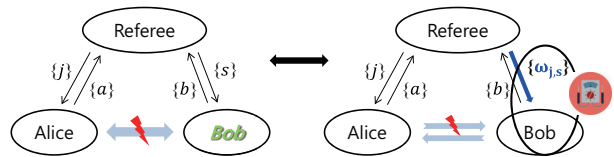}
\caption{(Left) In the EPR steering scenario, the referee provides classical information to and receives it from Alice and Bob. According to the payoff set by the referee, the total score of Alice and Bob is given by Eq.~(1). In this scenario, the referee trusts Bob, and no communication between Alice and Bob is allowed. We highlighted Bob's name with a bold Italic letter to denote the trust of the referee. (Right) The QRS game is a kind of variation of the Bell game wherein the transmission of information from the referee to Bob is given by linearly dependent quantum states. Bob performs joint measurements on his part of the shared pair and state given by the referee (denoted by an ellipse with a measurement apparatus in the figure). Bob cannot always determine with certainty which information is provided to him, thus the trust on Bob is removed and one-way communication from Bob to Alice may be permitted. Their total score is again given by Eq.~(1).}
\label{Fig1(Color Online)}
\end{figure}

  \section{Loss tolerant QRS game}
 \subsection{Settings for Loss Tolerant Scheme}

  In Ref.~\cite{Kocsis15}, a canonical way to convert a steering inequality to a corresponding score function was proposed. Nervertheless, Ref.~\cite{Kocsis15} assumed that no communication between Alice and Bob is allowed, and did not consider Bob's heralding efficiency. However, one-way communication from Bob to Alice is one of the many potential advantages of the QRS game, and coping with Bob's heralding efficiency is essential for the MDI scenario. Therefore, in order to make the QRS game MDI, and to fully exploit the advantages of it, we should consider one-way communication from Bob to Alice and the effect of heralding efficiency of Bob.

  There are two common ways of dealing with losses without post-selection. One way allows the experimentalist to report null results, say $\varnothing$, and investigate the effect. If we denote the set of \textit{valid} measurement outcomes as $\{a_i\}_i$, the set of \textit{all} measurement results becomes $\{a_i\}_i \cup \{\varnothing\}$ within this scheme. In steering verification, the effect of null results reported by parties changes the observed correlation between them. Hence, the corresponding steering bound should be reformulated. For the case where Alice is allowed to report null results, an $\eta_\textsc{h}$-dependent steering bound is constructed for two-qubit~\cite{Bennet12} and two-qudit systems~\cite{Skrzypczyk15}, where $\eta_\textsc{h}$ denotes the heralding efficiency of Alice. On the other hand, another way of dealing with losses is to do not accommodate the null results $\varnothing$ in the set of all measurement results. Therefore in each measurement, regardless of the occurrence of losses, the experimentalist has to choose one of the outcomes from the set $\{a_i\}_i$ to report. In the QRS game, since the goal of the parties is to maximize their score, their choices will be corresponding optimal values. These two ways are named in Ref.~\cite{Evans13} as $\textquoteleft$Depression' and $\textquoteleft$Anger', respectively, and well analyzed in Refs.~\cite{Evans13, Evans14}

In this paper, we will use both strategies in dealing with losses - $\textquoteleft$Depression' for Alice and $\textquoteleft$Anger' for Bob. We chose the $\textquoteleft$Depression' scheme for Alice because we are going to utilize the result in Ref.~\cite{Bennet12}, which adopts $\textquoteleft$Depression' scheme, to derive QRS score function. Meanwhile, Bob's losses should be coped with using the $\textquoteleft$Anger' scheme, to prevent more chance of deceiving the referee using additional option $\varnothing$ and one-way communication from Bob to Alice. Furthermore, in the QRS score function, Bob's measurement outcome is one of $\{0,1\}$, and it will become clear from the form of the payoff function that they correspond to measurement success and failure, thus in our paper, null results are included in one of the binary results, $\textquoteleft$0'. This strategy may be generalized for any QRS game as it is also implied for some entanglement verification results~\cite{Branciard13,Verbanis16}.

We note that for the $\textquoteleft$Anger' scheme, the heralding efficiency is fixed to unity while this may not be the case for actual measurement efficiency. Therefore to analyze Bob's losses, we shall use the term measurement efficiency, while the term heralding efficiency will be used to indicate that of Alice.

\subsection{Score function}

   In this subsection, we convert the steering inequality obtained in Ref.~\cite{Bennet12},

 \begin{align}
 \frac{1}{n}\sum_{j=1}^{n}\langle a_j \hat{B}_j \rangle \leq C_n(\eta_\textsc{h}),
 \end{align}
  into the QRS score function, where $a_j$ is Alice's reporting value from $\{+1,-1\}$, $\hat{B}_j$ is Bob's dichotomic measurement such that its outcomes is one of $\{+1,-1\}$, $\eta_\textsc{h}$ is Alice's heralding efficiency, $n$ is a number of measurement settings, and $C_n(\eta_\textsc{h})$ is $\eta_\textsc{h}$-dependent steering bound. We can rewrite it as

 \begin{align}
 \frac{1}{n} \sum_{j=1}^{n} (\langle a_j \hat{B}_j \rangle  - C_n(\eta_\textsc{h}) )\leq 0.
 \end{align}

 This form can be considered as a score in the 1s-DI steering scenario such that they obtain $1-C_n(\eta_\textsc{h})$ as a payoff if Alice guess Bob's outcome correctly, and lose $1+C_n(\eta_\textsc{h})$ otherwise. In the QRS game, it is the information $s$ provided by the referee that both Alice and Bob have to guess, thus $\hat{B}_j$ should be replaced by $s$. Moreover, the measurement efficiency of Bob does not appear here because in 1s-DI steering, we trust Bob, so if Bob reports nothing, then we conclude that a round has not started. This should also be inserted in the QRS score function as a factor $b \in \{0,1\}$ indicating measurement success or failure. The total payoff in the QRS game is then converted to

 \begin{align} \label{payoff}
 \mathcal {P} (a, b, j, s) = ( as - C_n(\eta_\textsc{h}) )b.
 \end{align}
 To construct the score function, it is optimal to prepare $j\in \mathcal{J}$ and $s \in \mathcal{S}$ with equal probability distribution, respectively, because if there exists some bias, Alice and Bob may take advantage of this fact. In this case $\mathcal{J}=\{1, 2, ... , n\}$ and $\mathcal{S}= \{+1,-1\}$ so that $j$ takes same role in inequality (3) and $s$ takes over the role of outcome of $\hat{B}_j$ in inequality (3). Hence, setting $p(j)=\frac{1}{n}$ and $q(s)=\frac{1}{2}$ turns the score (1) in this case to

 \begin{align} \label{score}
S_n(\eta_\textsc{h}) = \frac{1}{2n}\sum_{j,s}[s \langle ab \rangle _{j,s} - C_n(\eta_\textsc{h}) \langle b \rangle_{j,s}],
 \end{align}
where we invoke $S_n(\eta_\textsc{h})$ to make it clear that the score function depends on a number of measurement settings and heralding efficiency of Alice. As a last step, we encode information $s$ in linearly dependent quantum states

  \begin{align} \label{quantum state}
 \omega_{j,s} = \frac{\hat{I}_2 + s \hat{B}_j}{2},
  \end{align}
  following Refs.~\cite{Branciard13, Kocsis15}. We note that the score function (\ref{score}) is not a definitive form because we have to analyze the effect of imperfect preparation of quantum states (\ref{quantum state}) by the referee. We will revisit this problem at the end of this subsection and derive the definitive form in Eq. (\ref{score r})

In order to prove that the score function is indeed a steering criterion even when one-way communication is allowed from Bob to Alice, we will show that any unsteerable state cannot yield a positive score. Recall that an unsteerable state is such that its measurement outcomes can be explained by a combination of local realistic theory and local quantum model. Therefore if Bob can send messages to Alice, the most general strategy is to perform some POVM on $\omega_{j,s}$ to guess the information $s$ and, next, send the guessed value, say $\overline{s}$, to Alice. Note that the dependence of distribution on some hidden variable $\lambda$ is removed in this process thanks to linearity. As a consequence, Alice's choice $a$ is determined by $j$ and $\overline{s}$. Since the payoff (4) for each round has no explicit $j$ dependence, optimal choice of Alice depends only on $\overline{s}$. This does not mean that $a$ is independent of $j$, rather, it implies that $a$ depends on $j$ implicitly \textit{via} $\overline{s}$. With the gained information $\overline{s}$, the deterministic optimal choice of Alice is to report $a=\overline{s}$ for favorable $j$'s which can contribute to elevate the score, and report $\varnothing$ otherwise. Let us denote the set of favorable $j$'s as $\mathcal{F}_J$ and $|\mathcal{F}_J|=F$. Then the total score reads

\begin{align}
 S = \frac{\gamma}{2F}\sum_{j\in \mathcal{F}_J,s,\overline{s}}(\overline{s}\,s - C_n(\eta_\textsc{h}))\,p(\overline{s}|j,s) ,
 \end{align}
where $\gamma$ is the relative frequency that Bob reports $1$. Let us denote $+1$ as $+$ and $-1$ as $-$ for simplicity. Then using the fact that $p(+|j,s)+p(-|j,s) =1$, summing over $s$ and $\overline{s}$ gives

\begin{align}
 S = \frac{\gamma}{F}\sum_{j\in \mathcal{F}_J}(p(+|j,+) + p(-|j,-) -1 - C_n(\eta_\textsc{h})).
 \end{align}

To calculate Eq.~(8), let us write $\hat{B}_j = \vec{b}_j\cdot \hat{\vec{\sigma}}$ where $\vec{b}_j$ is a three dimensional vector whose norm is less than or equal to one, and $\hat{\vec{\sigma}}$ is a pseudovector of Pauli operators $(\hat{\sigma}_x, \hat{\sigma}_y, \hat{\sigma}_z)$. Any POVM element on two qubit system can be written as $\mu(\hat{I}_2 + \vec{m}\cdot \hat{\vec{\sigma}})$ where $\vec{m}$ is again a three dimensional vector whose norm is less than or equal to one, and $\mu$ satisfies $0\leq \mu\leq \frac{1}{1+|\vec{m}|}$. We then have

\begin{align} \nonumber
 p(+|j,+)&=\frac{1}{2}\mathrm{Tr}[\mu(\hat{I}_2+\vec{m}\cdot \hat{\vec{\sigma}})(\hat{I}_2+\vec{b}_j\cdot \hat{\vec{\sigma}})]\\ \nonumber
&= \mu(1+\vec{m}\cdot\vec{b}_j),\\ \nonumber
 p(-|j,-)&=\frac{1}{2}\mathrm{Tr}[((1-\mu)\hat{I}_2 - \mu \vec{m}\cdot \hat{\vec{\sigma}})(I_2-\vec{b}_j\cdot \hat{\vec{\sigma}})]\\
 &=1-\mu+\mu \vec{m}\cdot\vec{b}_j.
 \end{align}
Therefore, the score reads

\begin{align}
  S = \frac{\gamma}{F}\sum_{j\in \mathcal{F}_J}(2 \mu \vec{m}\cdot\vec{b}_j - C_n(\eta_\textsc{h})).
 \end{align}
First, we observe the inequality

\begin{align}
\sum_{j\in \mathcal{F}_J} \frac{2}{F} \mu \vec{m}\cdot\vec{b}_j \leq \frac{2 |\vec{m}|}{1+|\vec{m}|}\left |\frac{\sum_{j}\vec{b}_j}{F} \right |\nonumber \leq \left |\frac{\sum_{j}\vec{b}_j}{F} \right |,
\end{align}
where the first inequality is obtained by maximizing $\mu$ and using Cauchy's inequality, and the second inequality originates from the condition that $|\vec{m}|$ is less than or equal to one. Using the fact that the positive eigenvalue of the $\vec{v} \cdot \hat{\vec{\sigma}}$ is $|\vec{v}|$, we have

 \begin{align}
  \left |\frac{\sum_{j}\vec{b}_j}{F} \right | = \lambda_{max} \left [\frac{\sum_{j}(\vec{b}_j \cdot \hat{\vec{\sigma}})}{F} \right ] = \lambda_{max} \left[ \frac{1}{F} \sum_{j\in \mathcal{F}_J} \hat{B}_j \right],
 \end{align}
where $\lambda_{max}$ denotes the maximum eigenvalue of the argument operator. It is obvious that if we let the coefficients of each $\hat{B}_j$ be one of $+1$ or $-1$, then the maximization over such coefficients will bound Eq.~(11) from above. Thus, we have

 \begin{align}
  \lambda_{max} \left[ \frac{1}{F} \sum_{j\in \mathcal{F}_J} \hat{B}_j \right] \leq \max_{\{A_j\}_m} \left \{ \lambda_{max} \left[ \frac{1}{F} \sum_{j\in \mathcal{F}_J} A_j \hat{B}_j \right] \right \},
 \end{align}
where $A_j \in \{ +1, -1 \}$ for each $j \in \mathcal{F}_J$. One can see that the righthand side of inequality (12) is exactly the expression of $D_n(\eta_\textsc{H})$ in Eq.~(3) in Ref.~\cite{Bennet12}. Therefore if we generalize Alice's optimal strategy to probabilistic mixtures of deterministic choices, the righthand side of inequality (12) saturates at $C_n(\eta_\textsc{h})$. For more details of probabilistically mixing optimal deterministic strategies, see Sec. II-B in Ref.~\cite{Bennet12}. As a consequence, we obtain a bound

\begin{align}\label{final}
 \max_{\{w_k\}}\left [ \sum_{k=1}^{n} \frac{k}{n \eta_\textsc{h}} w_k \left ( \sum_{j\in \mathcal{F}_J} \frac{2}{F} \mu \vec{m}\cdot\vec{b}_j \right ) \right ] \leq C_n(\eta_\textsc{h}),
\end{align}
where $w_k$ satisfies $0 \leq w_k \leq 1$ and $\sum_{k=1}^{n} w_k=1$. When inequality (12) is saturated, inequality (\ref{final}) is also saturated according to the definition of $C_n(\eta_\textsc{h})$ (see, Eq.(4) in Ref.~\cite{Bennet12}), which shows that $C_n(\eta_\textsc{h})$ is a tight bound for inequality (\ref{final}).  This concludes that the score when Bob reports $1$ is bounded from above by $0$ even we allow one-way communication from Bob to Alice. Therefore, it is impossible to get a  positive score using unsteerable states, and consequently, the score function (\ref{score}) is indeed a steering criterion.

We note here that the strategy can be optimized once more. Bob can report $1$ with some weight according to his measurement outcome. However, the effect of a weighted report is compromised by reporting $a=\overline{s}$. This is apparent from the form of the payoff (\ref{payoff}), because Alice and Bob will consider $\overline{s}$ to be $s$, and think that they can best adjust $as$ term in the payoff (\ref{payoff}) to $1$ by reporting $a=\overline{s}$. Therefore the payoff they expect to obtain is symmetric in $\overline{s}$, which removes the effect of weighted report and does not increase the score for unsteerable states.


The foregoing argument is developed under the assumption that the quantum state $\omega_{j,s}$ provided by the referee is perfectly prepared in the form of Eq.~(\ref{quantum state}). In Ref.~\cite{Kocsis15}, however, this assumption is removed by analyzing the effect of imperfect preparation of the state $\omega_{j,s}$ and introducing the factor $r$ multiplied to the steering bound $C_n(\eta_\textsc{h})$ to compensates the imperfection. If the referee fails to prepare $\omega_{j,s}$ and some state appears frequently as a result, untrusted parties can exploit this imperfection to maximize their score. As an extreme example, if the referee prepares $\omega_{j,s}$ as $\frac{\hat{I}_2 + s \hat{B}_1}{2}$ and Bob performs POVM $\hat{B}_1$, Bob can always determine $s$ with certainty, thus by reporting $a=s$ and $b=1$, they obtain the optimal score $1-C_n(\eta_\textsc{h})$. Therefore we need to suppress the undesired elevation of the score, which can be accomplished by adding a factor $r$ in front of the steering bound

\begin{align}\label{score r}
S_n(\eta_\textsc{h}, r) = \frac{1}{2n}\sum_{j,s}[s \langle ab \rangle _{j,s} - r C_n(\eta_\textsc{h}) \langle b \rangle_{j,s}],
\end{align}
where we include $r$ as an argument of the score function. This is the definitive form of the score function. The detailed method to obtain $r$ is presented in the Methods section in Ref.~\cite{Kocsis15}. For our score function, assuming that the referee prepares $\omega_{j,s}$ as $\frac{\hat{I}_2+ \vec{n}_{j,s} \cdot \hat{\vec{\sigma}}}{2}$, the factor $r$ is calculated to be

\begin{align} \label{r-value}
  r = \max_{\{a_j=\pm 1 \}} \frac{- \langle \vec{A}, \vec{B} \rangle + \sqrt{\langle \vec{A}, \vec{B} \rangle ^2 + \langle \vec{A}, \vec{A} \rangle (n^2-\langle \vec{B}, \vec{B} \rangle)}}{C_n(\eta_\textsc{h}) (n^2-\langle \vec{B}, \vec{B} \rangle)},
\end{align}
where $\vec{A} = \sum_{j} a_j \frac{(\vec{n}_{j+}-\vec{n}_{j-})}{2}$, $\vec{B} = \sum_{j}\frac{(\vec{n}_{j+}+\vec{n}_{j-})}{2}$.
Unfortunately, introducing the factor $r$, however, cannot fully remove the existence of trust. This is because we have to perform tomography on the state provided by the referee to obtain $\vec{n}_{j,s}$, and tomography requires trust on measurement devices. Nonetheless, as explained in Sec. I, the QRS game is more reliable verification than 1s-DI steering.


 Now let us consider the case where Bob suffers from losses. It is obvious that the optimal strategy for Bob when losses occur is to perform some POVM on the state provided by the referee, since he has no access to the complete bipartite system. Indeed, this strategy is identical to the case of sharing unsteerable states, such that the maximal score is bounded from above by 0. Therefore the optimal strategy of Bob to deal with losses is to report $0$, which results in the shrinkage of the score by measurement efficiency, say, $\eta_\textsc{m}$. That is, if we denote the score by $S$ when the measurement efficiency of Bob is perfect, losses reduce the score to $\eta_\textsc{m} S$. It is obvious that multiplying by $\eta_\textsc{m}$ does not change the sign of the score unless $\eta_\textsc{m}$ is zero, thus non-zero measurement efficiency of Bob does not affect the steerability at all. This guarantees arbitrarily loss-tolerant verification of steering with the result in Ref.~\cite{Bennet12} that inequality (2) can be violated using a maximally entangled state if a number of measurement settings is larger than the reciprocal of the heralding efficiency of Alice, say, $n > 1/\eta_\textsc{h}$.

 The loss-tolerant property of our QRS game is asymmetric with respect to Alice and Bob. While the measurement efficiency of Bob does not affect steerability, the heralding efficiency of Alice is reflected in the steering bound $C_n(\eta_\textsc{h})$ and can change the sign of the score function for given quantum states and a number of measurement settings. This property corresponds to that of a loss-tolerant 1s-DI steering scenario in which only the heralding efficiency of Alice is of concern, while the measurement efficiency of Bob is ignored by discarding experiments Bob failed to report. Regardless, it is a newly found asymmetry property for untrusted steering parties which was previously not observed before to our knowledge. We belive that this property can be applied to practical asymmetric quantum information tasks.

\section{Conclusion}
In this study, we have converted the arbitrarily loss-tolerant steering inequality in Ref.~\cite{Bennet12} to the corresponding score function (\ref{score r}) in the QRS game, and showed that it is arbitrarily loss-tolerant when both parties cannot be trusted. To do this, we permitted Alice to report null results while Bob is prohibited to do so. We showed that our score function is indeed a steering criterion; unsteerable states cannot obtain positive value, even with the help of one-way communication from Bob to Alice. Moreover, to compensate the effect of imperfect state preparation, we introduced and calculated a closed form of factor $r$ in (\ref{r-value}) to suppress any undesired elevation of the score using unsteerable states.

We recapitulate that the verification of steering depends on the score function, that is, some state $\rho_{AB}$ can be determined as a steerable state by a score function $S_n(\eta_\textsc{h}, r)$, while it is determined as an unsteerable state by another score function $S_{n\textquoteright}(\eta_\textsc{h} ', r\textquoteright)$. However, the verifiability of steering does not depend on the measurement efficiency of Bob unless it is zero. This reveals additional asymmetry property of the steering verification.

The MDI characteristic broadens the application scope of quantum information tasks that were not previously possible, such as unconditionally secure communication, and the loss-tolerant property allows one to implement such tasks in a realistic environment. Furthermore, the asymmetry property found here can be used for asymmetric information tasks so that only one party is free from the threat of losses. We thus expect that our work is relevant for realizing useful and practical quantum information tasks.

\section{Acknowledgments}
The authors thank Wonmin Son, Yong-Siah Teo and Hyukjoon Kwon for useful discussions and comments. This work was supported by a National Research Foundation of Korea grant funded by the Korea government (MSIP) (No. 2010-0018295) and by the KIST Institutional Program (Project No. 2E27800-18-P043).

\end{document}